\documentclass[aps,prb,floatfix,twocolumn,notitlepage,superscriptaddress,10pt]{revtex4-2}
\usepackage{xcolor}
\usepackage{grffile}
\usepackage{amsmath, amsthm, amssymb, bbold}
\usepackage[normalem]{ulem}
\usepackage{cancel}
\usepackage{bm}
\usepackage{microtype}
\usepackage{hyperref}
\usepackage{setspace}
\usepackage{grffile}
\hypersetup{colorlinks,linkcolor=blue,urlcolor=blue,citecolor=blue}
\usepackage{amsmath}    
\usepackage{amssymb}
\usepackage{graphicx}   
\usepackage{verbatim}   
\usepackage{color}      
\usepackage{microtype}
\usepackage[normalem]{ulem}
\usepackage{natbib}
\usepackage{enumitem} 
\usepackage{dsfont} 
\usepackage{hyperref}   

\newcommand{\ket}[1]{|#1\rangle}
\newcommand{\average}[1]{\langle #1\rangle}

\newcommand{\dbra}[1]{\langle\!\langle #1}
\newcommand{\dket}[1]{|#1\rangle\!\rangle}

\begin{document}

\title{Hot but Coherent: Doublons at Infinite Temperature in the Hubbard chain}

\author{C\u at\u alin Pa\c scu Moca}
\email{mocap@uoradea.ro}
\affiliation{Department of Physics, University of Oradea,  410087, Oradea, Romania}
 \affiliation{Department of Theoretical Physics, Institute of Physics, Budapest University of Technology and Economics, M\H{u}egyetem rkp.~3, H-1111 Budapest, Hungary}
\author{Bal\'azs D\'ora}
\affiliation{Department of Theoretical Physics, Institute of Physics, Budapest University of Technology and Economics, M\H{u}egyetem rkp.~3, H-1111 Budapest, Hungary}
\author{Gergely Zar\'and}
\affiliation{Department of Theoretical Physics, Institute of Physics, Budapest University of Technology and Economics, M\H{u}egyetem rkp.~3, H-1111 Budapest, Hungary}

\date{\today}

\begin{abstract}
We investigate the spectral properties and the dynamics of doublons in the 
one-dimensional Hubbard model at infinite temperature. Using a Chebyshev expansion approach formulated in the 
superfermionic 
representation, we compute the momentum- and frequency-resolved doublon spectral function across a wide range of 
interaction 
strengths $U$ and in the presence of an external electric field. Increasing the interaction, the spectrum 
gradually splits into separate bands 
associated with two-hole and two-particle excitations. Despite the presence of strong correlations, we find that 
doublons retain their coherence and undergo long-lived Bloch oscillations, as well as rich quantum walk dynamics 
characterized by light-cone spreadings at strong coupling, which we analyze through the time-and space-resolved Green's function.
\end{abstract}

\maketitle

\section{Introduction}

Understanding the nature of excitations in strongly correlated quantum systems is a central challenge in condensed matter physics. Among the most 
intriguing and experimentally relevant excitations in the Hubbard model are doublons—bound states of two fermions occupying the same lattice 
site. These composite objects arise naturally when a finite on-site repulsion $U$ is present, and they play a crucial role in the non-equilibrium 
dynamics and transport properties of Mott insulators, cold-atom systems, and correlated electron materials~\cite{jaksch1998cold,strohmaier2010observation, sensarma2010dynamics}. Despite their large formation energy, typically of order $U$, doublons can display unexpectedly long 
lifetimes and coherent motion, especially in low-dimensional systems. Their stability has been attributed to energy constraints and dynamical 
selection rules that hinder decay into lower-energy states~\cite{sensarma2010dynamics, eckstein2011thermalization}.

The spectral properties of doublons encode important information about their internal structure. In particular, the 
doublon spectral function $A_d(k,\omega)$ provides a momentum- and energy-resolved fingerprint of these excitations. 
 While the behavior of doublons at zero or low temperatures has been extensively studied~\cite
{heidrich2010doublon, rausch2017filling,sensarma2010dynamics,heidrich2010doublon,Hofmann_2012,Rausch_2016,benthien2004spectral,benthien2007spin,carmelo2006general}, much less is known about their fate in the high-temperature limit, where strong thermal fluctuations 
are expected to suppress quasiparticle coherence. In the extreme case of infinite temperature, where the system is in a maximally mixed state, it 
is unclear whether doublons remain meaningful or whether their spectral signatures are washed out entirely~\cite{nocera2018finite,tindall2019heating}.

In this work, we address this question by systematically studying the spectral and dynamical properties of doublons in the one-dimensional 
Hubbard model at infinite temperature. Our analysis is based on a tensor network approach using matrix product states (MPS)~\cite{schollwock2011density, paeckel2019time}, adapted to simulate the vectorized density matrix in the so-called superoperator formalism~\cite{dzhioev2011super, kshetrimayum2019tensor}. 
Within this framework, we compute spectral functions using an adapted Chebyshev expansion method~\cite{weisse2006kernel, braun2014numerical, wolf2014chebyshev, ganahl2014chebyshev} which is well-suited for high-resolution calculations free from 
finite-time truncation effects.

By working in the infinite-temperature regime—where thermal smearing is maximal—we can disentangle interaction-driven effects from those arising purely due to temperature. Surprisingly, we find that well-defined doublonic features persist even at infinite temperature, 
both in the spectral domain and in real-time dynamics. When a static electric field is applied, we observe the formation of a Wannier-Stark 
ladder~\cite{mendez1993wannier} in the spectral response, and detect robust Bloch oscillations~\cite{dahan1996bloch,hartmann2004dynamics} in the doublon propagator—signatures of coherent transport that survive 
strong thermal disorder. As the interaction strength is increased, the doublon spectrum splits into distinct two-hole and two-particle bands, 
revealing how interaction and field effects compete in shaping doublon dynamics.

Beyond their spectral properties, doublons also exhibit rich real-time dynamics when locally injected into an 
infinite-temperature background. Such doublonic quantum walks reveal distinctive light-cone structures governed 
by both free-particle and interaction-renormalized propagation velocities~\cite{ostahie2023multiparticle,giri2021two,paul2024realizing}.  Although doublon number is not conserved due to its 
non-commutativity with the kinetic part of the Hamiltonian, we observe that its decay is progressively suppressed with 
increasing interaction strength. 
When a static electric field is applied, the system exhibits clear Bloch oscillations in the doublon 
density even at infinite temperature, indicating the persistence of coherent doublon motion while  they 
become gradually damped as interactions are turned on. 

\section{Theoretical framework}

The Hubbard model~\cite{hubbard1963electron,essler2005one,arovas2022hubbard} is a fundamental lattice model used to describe the interplay between kinetic energy and local 
electron-electron interactions in correlated fermionic systems. The model consists of spinful fermions hopping between nearest-neighbor sites of a lattice, with an interaction term that penalizes double occupancy. The Hamiltonian reads
\begin{gather}
H = -{J\over 2} \sum_{i=-L/2}^{L/2} \sum_{\sigma = \uparrow, \downarrow} \left( c^\dagger_{i\sigma} c_{i+1,\sigma} + \text{h.c.} \right) \nonumber\\
	+	U \sum_{i=-L/2}^{L/2} \left(n_{i\uparrow}-{1\over 2}\right)\left(n_{i\downarrow}-{1\over 2}\right),
\label{eq:Hamiltonian}
\end{gather}
where $J$ denotes the hopping amplitude and $U$ the onsite interaction strength. We compute both the time-dependent 
Green’s function $G_d(x,t)$ and the momentum-resolved spectral function $A_d(k,\omega)$ associated with the doublonic dynamics, 
using an ensemble corresponding to infinite temperature, where all microstates are equally probable~\cite{moca2023kardar,penc2024loss}.

The doublon creation operator in real space is defined as $d_x^\dagger = c_{x\uparrow}^\dagger c_{x\downarrow}^\dagger$, 
which creates a local doublon by placing both a spin-up and a spin-down fermion on the same lattice site $x$. To probe 
the dynamical properties of such composite excitations, we consider the two-hole Green’s function—closely related to the 
lesser component of the Keldysh Green’s~\cite{freericks2019} function—given by
\begin{gather}
G_{2\text{-}h}(x,t) = -i \langle d^\dagger_{x}(t) d_0(0) \rangle_{\infty},
\label{eq:G_2h}
\end{gather}
where $\langle \cdot \rangle_{\infty}$ denotes averaging over the infinite-temperature ensemble, and $d_x(t)$ is the Heisenberg-evolved operator at time $t$.
To access momentum-and frequency-resolved information, we introduce the doublon operator in momentum space,
\begin{gather}
d_k^\dagger = \frac{1}{\sqrt{L}} \sum_{x} e^{ikx} d_x^\dagger = \frac{1}{\sqrt{L}} \sum_{q} c^\dagger_{k+q,\uparrow} c_{q,\downarrow},
\label{eq:dk}
\end{gather}
and define the corresponding momentum-resolved Green’s function as
\begin{gather}
G_{2\text{-}h}(k,t) = -i \langle d_k^\dagger(t) d_k(0) \rangle_{\infty}.
\end{gather}
The frequency-domain Green’s function is obtained via Fourier transformation:
\begin{gather}
G_{2\text{-}h}(k,\omega) = \int_{-\infty}^{\infty} dt e^{i\omega t} G_{2\text{-}h}(k,t).
\end{gather}

The particle (greater) contribution to the doublonic Green’s function, denoted by $G_{2\text{-}p}(k,\omega)$, is computed analogously by 
exchanging the roles of creation and annihilation operators in the time-ordered correlator, i.e., $d_k^\dagger \leftrightarrow d_k$. This 
symmetry reflects the underlying particle-hole structure of the doublon dynamics and ensures that both creation and annihilation processes 
are fully incorporated.
The total doublon spectral function is defined as the imaginary part of the corresponding retarded Green’s function:
\begin{gather}
A_d(k,\omega) = -2 \mathrm{Im}  G_d(k,\omega),
\end{gather}
and serves as a measure of the density of doublon states at energy $\omega$ and momentum $k$. It consists of both particle-like and hole-like contributions:
\begin{gather}
A_d(k,\omega) = A_{2\text{-}p}(k,\omega) + A_{2\text{-}h}(k,\omega),
\end{gather}
providing a complete spectral fingerprint of doublon excitations in the system.

An important quantity is the momentum-integrated spectral function, which provides a global view of the doublon density of states across all momenta. It is defined as
\begin{equation}
A_d(\omega) = \frac{2\pi}{L} \sum_k A_d(k,\omega),
\label{eq:A_int}
\end{equation}
This quantity is particularly useful for analyzing the total spectral weight distribution and identifying high-energy features associated with 
doublon excitations. In contrast to the momentum-resolved function $A_d(k,\omega)$, which highlights the dispersion and coherence of doublons, 
the integrated spectral function emphasizes their energy distribution and is less sensitive to finite-size effects or the details of 
translational symmetry breaking.

To investigate the response of doublonic excitations to external driving, we consider the effect of a static electric field applied to the Hubbard 
model. There are two common gauge choices for incorporating such a field on the lattice~\cite{Davies.1988}. In the scalar potential gauge, the electric 
field enters as a site-dependent linear potential, which explicitly breaks translational symmetry and modifies the Hamiltonian as
\begin{equation}
H_E = H + E \sum_x x\,  n_x,
\end{equation}
where E denotes the electric field strength and $n_x = n_{x\uparrow} + n_{x\downarrow}$ is the local particle number operator. This formulation 
directly tilts the potential landscape, making the localization and energy shift of doublon excitations, at the cost of losing momentum as a good quantum number.

Alternatively, in the vector potential gauge, the electric field is introduced via a time-dependent vector potential $A(t)$, which enters the hopping 
terms through a Peierls substitution: $t \rightarrow t\, e^{iA(t)}$ in Eq.~\eqref{eq:Hamiltonian}. For a static electric field E, this corresponds to a 
linearly growing vector potential $A(t) = -Et$. The Hamiltonian in this gauge becomes time-dependent~\cite{clade2017improving}.
This approach preserves translational invariance, allowing for a momentum-resolved description of the doublon dynamics, though the time dependence complicates both analytical and numerical treatments.

In either gauge, the electric field effectively tilts the energy landscape and induces Bloch oscillations with a characteristic frequency $\omega_B 
\propto eE/\hbar $\cite{Eckstein.2011,Rausch_2016}. These oscillations significantly affect the propagation and coherence of doublons, leading to 
shifted spectral weight and—in some regimes—dynamical localization\cite{rausch2017filling,Hofmann_2012}. In what follows, we adopt natural units with $e = \hbar = 1$, so the Bloch frequency simplifies to $\omega_B = E$.

\subsection{Numerical approach}
To compute the spectral function directly in frequency space, we use the 
Chebyshev polynomial expansion method adapted to work with density matrices~\cite{cui2015variational,jaschke2018one,casagrande2021analysis,nakano2021tensor}. 
This approach circumvents explicit time evolution 
and instead approximates spectral quantities via recursive polynomial constructions, 
making it particularly effective in systems with strong interactions.

A key step in this formulation is the vectorization of the density matrix, which allows us to recast expectation values as inner products in an 
enlarged Hilbert space. We introduce the dual set of fermionic operators, $\tilde{c}_{x\sigma}$ and $\tilde{c}^\dagger_{x\sigma}$, acting on 
a dual Fock space and obeying standard anticommutation relations: $\{\tilde{c}_{x\sigma}, \tilde{c}^\dagger_{x’\sigma’}\} = \delta_{x,x’}\delta_
{\sigma,\sigma’}$~\cite{dzhioev2011super,dzhioev2012nonequilibrium}. For a single site, the full Hilbert space becomes a tensor product of  regular and dual states,
\begin{equation}
\ket{\alpha, \tilde{\alpha}} = \left (\ket{0}, \ket{\uparrow}, \ket{\downarrow}, \ket{\uparrow\downarrow}\right ) \otimes 
 (\ket{\tilde{0}}, \ket{\tilde{\uparrow}}, \ket{\tilde{\downarrow}}, \ket{\tilde{\uparrow}\tilde{\downarrow}}).
\end{equation}
This “doubled” representation allows us to express the infinite-temperature density matrix $\rho \propto \mathbb{1}$ as a product state,
\begin{align}
\dket{\rho} &= \bigotimes_{j=-L/2}^{L/2} \exp\left(-i\sum_{\sigma} c^\dagger_{j\sigma} \tilde{c}^\dagger_{j\sigma}\right) \ket{0,\tilde{0}} \label{eq:GS}
&= \bigotimes_{j=-L/2}^{L/2} \dket{\rho_j},
\end{align}
where $\ket{0,\tilde{0}}$ is the vacuum in the doubled space. At half filling, the local vectorized state becomes~\cite{dzhioev2011super, dzhioev2012nonequilibrium}
\begin{gather}
\dket{\rho_j} = \frac{1}{4} \left( \ket{0,\tilde{0}}_j - i\ket{\uparrow,\tilde{\uparrow}}_j - i\ket{\downarrow,\tilde{\downarrow}}_j + \ket{\uparrow\downarrow,\tilde{\uparrow}\tilde{\downarrow}}_j \right).
\end{gather}
The state $\dket{\rho}$, as constructed in Eq.~\eqref{eq:GS} serves as a convenient reference state for computing expectation values. For instance, one finds $\langle n_{j\sigma} \rangle = \langle \tilde{n}_{j\sigma} \rangle = 1/2$, and the combined occupation ${\cal N}_{j\sigma} = n_{j\sigma} - \tilde{n}_{j\sigma}$ satisfies $\langle {\cal N}_{j\sigma} \rangle = 0$, consistent with an infinite-temperature ensemble.
The construction can be generalized to include a charge and spin imbalance  $\mu$ and $\mu_z$ via
\begin{gather}
\dket{\rho_j} = \frac{1}{Z} \left( \ket{0,\tilde{0}}_j - i\ket{\uparrow,\tilde{\uparrow}}_j e^{-\mu+\mu_z} \right.\nonumber \\
\left.- i\ket{\downarrow,\tilde{\downarrow}}_j e^{-\mu-\mu_z} + \ket{\uparrow\downarrow,\tilde{\uparrow}\tilde{\downarrow}}_j e^{-2\mu} \right),
\end{gather}
with $Z$ the normalization constant, allowing access to infinite temperature density matrix away from half-filling. 
\begin{figure*}[htb!]
	\begin{center}
	 \includegraphics[width=1.9\columnwidth]{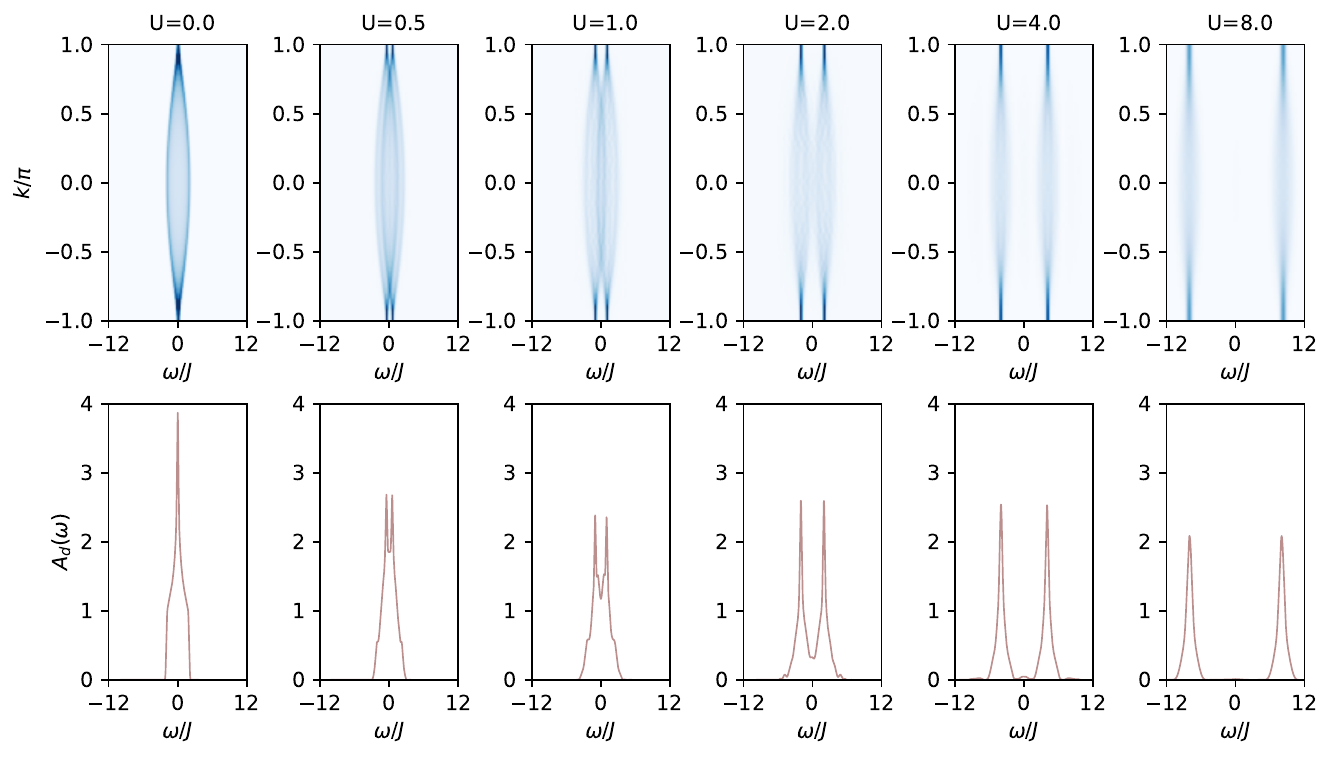}
	 \caption{Top row: Momentum-resolved doublon spectral function $A_d(k,\omega)$ for various interaction strengths $U$, as labeled in each panel. The results are obtained using the Chebyshev expansion method for a system of size $L=50$.
Bottom row: Corresponding momentum-integrated spectral functions, as defined in Eq.~\eqref{eq:A_int}. For $U=0$, the spectrum consists of a single continuous band confined to the interval $\omega \in [-2J, 2J]$, featuring a prominent resonance at $\omega = 0$ arising from contributions near momenta $k \approx \pm \pi$. As the interaction strength $U$ increases, the spectral weight gradually redistributes and begins to separate. In the strong-coupling regime, the spectrum clearly splits into two well-defined bands: a lower-frequency band dominated by two-hole ($2\text{-}h$) excitations, and an upper-frequency band associated with two-particle ($2\text{-}p$) processes.}
	 \label{fig:Ad_k_w}
	\end{center}
\end{figure*}
To compute the spectral function directly in frequency and momentum space, we employ an extension of the Chebyshev polynomial expansion method adapted to the vectorized density matrix formalism~\cite{holzner2011chebyshev,ganahl2014chebyshev}. In this framework, the imaginary part of the Green’s function is expanded in terms of Chebyshev polynomials $T_n(\mathcal{L})$ of the rescaled Liouvillian superoperator,
\begin{equation}
\mathcal{L} = H \otimes \mathbb{1} - \mathbb{1} \otimes H,
\label{eq:L}
\end{equation}
which governs the dynamics in Liouville space. Since Chebyshev polynomials are defined on the interval $[-1, 1]$, the spectrum of $\mathcal{L}$ must be linearly rescaled to lie within this domain.
To perform the rescaling, we first determine the spectral boundaries $L_{\text{min}}$ and $L_{\text{max}}$ of $\mathcal{L}$ via two separate DMRG runs. We then apply a linear transformation using the parameters
\begin{equation}
a = \frac{L_{\text{max}} - L_{\text{min}}}{2 - \delta}, \qquad b = \frac{L_{\text{max}} + L_{\text{min}}}{2},
\end{equation}
where $\delta \sim 0.01$ ensures that the rescaled spectrum remains strictly within $[-1,1]$, maintaining numerical stability in the Chebyshev recursion. This rescaling is also applied to the physical frequencies, such that $\omega = a \omega' + b$, where $\omega' \in [-1, 1]$.

Using this setup, the momentum-resolved spectral function corresponding to two-hole excitations is approximated by the Chebyshev expansion:
\begin{equation}
A_{2\text{-}h}(k,\omega) \approx \frac{1}{a} \frac{1}{\sqrt{1 - \omega'^2}} \left[ \mu_0 + 2 \sum_{n=1}^{N_{\text{max}}} \mu_n T_n(\omega') \right],
\label{eq:A_2h_ext}
\end{equation}
where the Chebyshev moments $\mu_n$ encode the system’s dynamical response and are computed through recursive applications of $\mathcal{L}$ on the initial vectorized state.
\begin{equation}
\mu_n = \dbra{\rho}| (d^\dagger_{k} \otimes \mathbb{1}) T_n({\cal L}) (d_{k} \otimes \mathbb{1}) \dket{\rho},
\label{eq:mu_2h}
\end{equation}
with $d^\dagger_{k}$ defined as in Eq.~\eqref{eq:dk}.
These moments are computed recursively via:
\begin{eqnarray}
\dket{b_0} &= &(d_{k} \otimes \mathbb{1}) \dket{\rho}, \nonumber \\
\dket{b_1} &= &\mathcal{L} \dket{b_0}, \\
\dket{b_{n+1}} &= &2 \mathcal{L} \dket{b_n} - \dket{b_{n-1}}, \nonumber
\end{eqnarray}
and $\mu_n = \dbra{b_0}  \dket{b_n}$. 
To suppress Gibbs oscillations from the finite expansion order $N_{\text{max}}$, 
we employ smoothing kernels such as the Jackson kernel~\cite{silver1996kernel,weisse2006kernel}, which improve convergence 
and spectral resolution. This makes the Chebyshev method ideally suited for computing the doublon spectral function $A_d(k,
\omega)$ with high accuracy across a broad frequency range. However, since the calculation is momentum-resolved, a 
separate recursion must be performed for each momentum $k$.

To explore doublonic Bloch oscillations on top of the infinite-temperature background, we compute the time- and position-resolved Green’s 
function $G_d(x,t)$ as defined in Eq.~\eqref{eq:G_2h}.
Time evolution under a Hamiltonian $H$ translates into a vectorized von-Neumann evolution
\begin{equation}
\dket{\rho(t)}= e^{-i\mathcal{L}t } \dket{\rho}.
\end{equation}
This allows us to compute the Green's function as
\begin{equation}
G_{2\text{-}h}(x,t) = -i \dbra{\rho}| (d^\dagger_{x} \otimes \mathbb{1}) 
e^{-i\mathcal{L}t} 
(d_{0} \otimes \mathbb{1}) \dket {\rho}.
\label{eq:G_d_vec}
\end{equation}
This formulation is particularly well-suited for MPS-based simulations~\cite{schollwock2013matrix}, where time evolution is implemented using Trotter-decomposed two-site 
gates acting on the doubled Hilbert space~\cite{vidal2007classical}. A key advantage of working at infinite temperature is that the initial vectorized state $\dket{\rho}$ 
is unentangled across the system–auxiliary bipartition. As a result, entanglement growth during the evolution remains minimal, allowing for 
efficient simulations with low computational overhead. All numerical simulations were performed using the ITensor library~\cite{fishman2022itensor}.
\begin{figure*}[htb!]
	\begin{center}
	 \includegraphics[width=1.9\columnwidth]{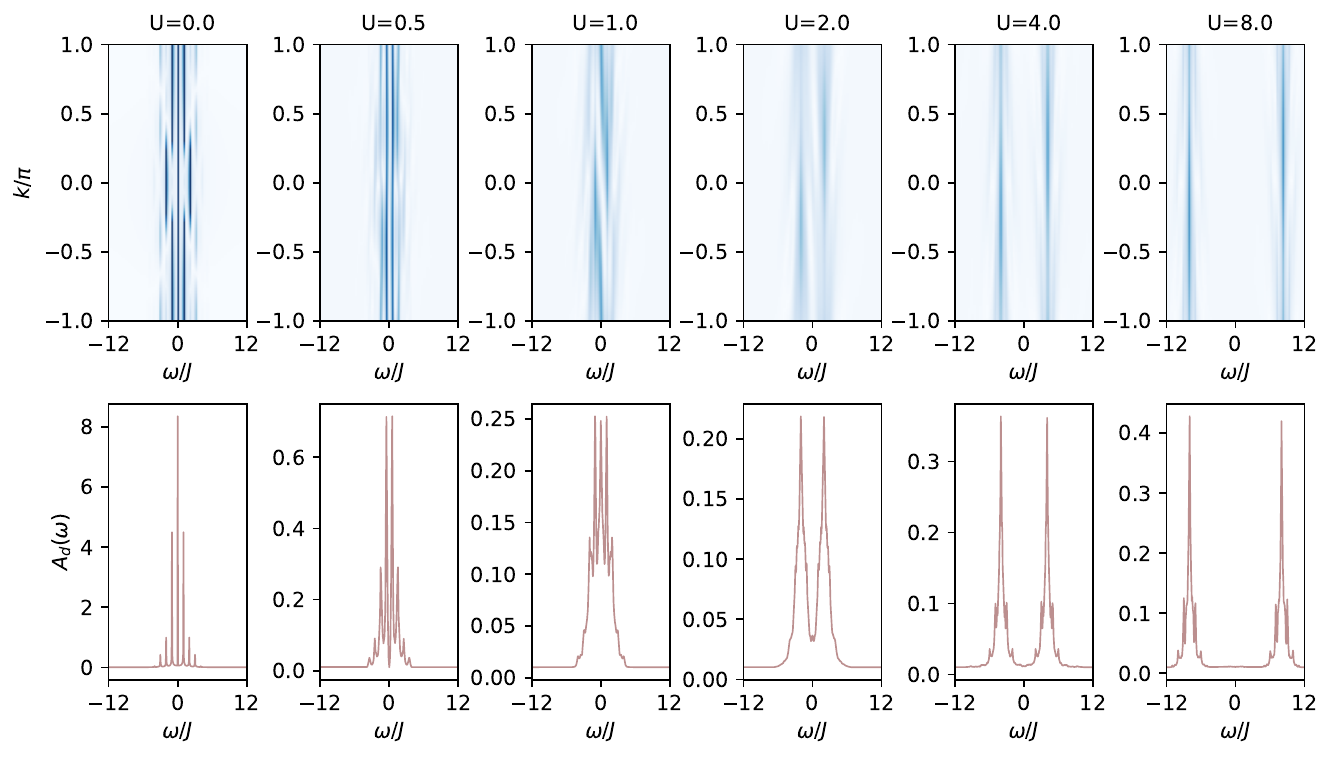}
	 \caption{
Top row: Momentum-resolved spectral function $A_d(k,\omega)$ computed for different interaction strengths $U$, as 
indicated in each panel, in the presence of an external electric field $E=1$.
Bottom row: Momentum-integrated spectral function, calculated using Eq.~\eqref{eq:A_int}. For $U=0$, the spectrum exhibits 
a Wannier–Stark ladder, with resonances located at integer multiples of the Bloch frequency $\omega_B$. As the interaction 
strength increases, the spectral weight gradually separates into two distinct bands corresponding to the two-hole ($2\text
{-}h$) and two-particle ($2\text{-}p$) sectors, similar to the $E=0$ case. Remarkably, the ladder-like resonances persist even at large values of $U$, indicating the robustness of the underlying Wannier–Stark structure. Here we used the vector potential gauge. }
	 \label{fig:Ad_E_k_w}
	\end{center}
\end{figure*}

\section{Spectral response and Bloch oscillations of doublons}

\subsection{Spectral functions}

In the non-interacting limit ($U=0$), the doublonic spectral function can be computed analytically by exploiting the quadratic nature of the Hamiltonian. 
Since doublons are composite objects formed by the product of two fermionic creation 
operators with opposite spins at the same site, their spectral properties at $U=0$ reduce to convolutions of single-particle 
propagators. In this regime, the system behaves as two independent spin species of free fermions, and Wick’s theorem applies~\cite{fetter2012quantum}. 
The doublon Green’s function, can thus 
be expressed as a product of spin-up and spin-down single-particle Green’s functions. After Fourier transforming to momentum 
and frequency space, the doublon spectral function $A_d(k,\omega)$ is obtained by integrating over internal momentum variables, 
constrained by total momentum conservation. 
The result reflects the joint density of states for two free fermions with opposite 
spin and total momentum $k$,
\begin{gather}
	A^{(0)}_d(k,\omega) = \frac{2\pi}{L}\sum_q \delta(\omega -\varepsilon_{k+q}-\varepsilon_{q}),
	\label{eq:A0d}
\end{gather} 
where $\varepsilon_k = -J \cos(k)$ denotes the single-particle energy dispersion and the superscript $(0)$ always stands 
for the $U=0$ limit. This result provides a  reference point for 
validating numerical methods and distinguishing features that arise purely from interactions at finite $U$. In one of the panels in Fig.~\ref
{fig:Ad_k_w}, we show the numerically obtained doublon spectral function $A^{(0)}_d(k,\omega)$ 
calculated using the Chebyshev expansion. As anticipated, at $k = \pm \pi$ the spectrum exhibits a sharp resonance at zero 
frequency, $A^{(0)}_d(k = \pm\pi, \omega) \propto  \delta(\omega)$, while for all momenta the spectral weight remains confined to the interval $\omega \in [-2J, 2J]$, consistent with the bounds of the two-particle continuum.

In the presence of a static electric field and in the non-interacting limit $U = 0$, the doublonic spectral function $A_d^{(0)}(\omega)$ exhibits the hallmark features of a Wannier-Stark ladder~\cite{Davies.1988}:
\begin{equation}
A_d^{(0)}(\omega) = \sum_n w_n\, \delta(\omega - n\omega_B),
\label{eq:WSL}
\end{equation}
where $\omega_B$ denotes the Bloch frequency, $w_n$ are the corresponding spectral weights, and $n \in \mathbb{Z}$ labels the ladder levels. The spectrum consists of delta-function peaks located at integer multiples of $\omega_B$, reflecting the quantized energy levels induced by the electric field.

The momentum-resolved spectral function $A_d^{(0)}(k,\omega)$ reveals analogous resonant structures, 
indicating coherent doublon dynamics in the presence of the field. To compute $A_d^{(0)}(k,\omega)$, we adopt the vector potential gauge, 
which preserves translational invariance and ensures that momentum $k$ remains a good quantum number.  However, it lacks time translational invariance therefore the doublon Green's function
in Eq. \eqref{eq:G_2h} should in principle depend on two time variables, corresponding to the doublon creation and annihilation operators. For the sake of simplicity, we fix the time variable
for the annihilation operator to zero and perform the temporal Fourier transform only for $t$.
These features are illustrated in one of the panels of Fig.~\ref{fig:Ad_E_k_w}.

\begin{figure*}[htb!]
	\begin{center}
	 \includegraphics[width=1.9\columnwidth]{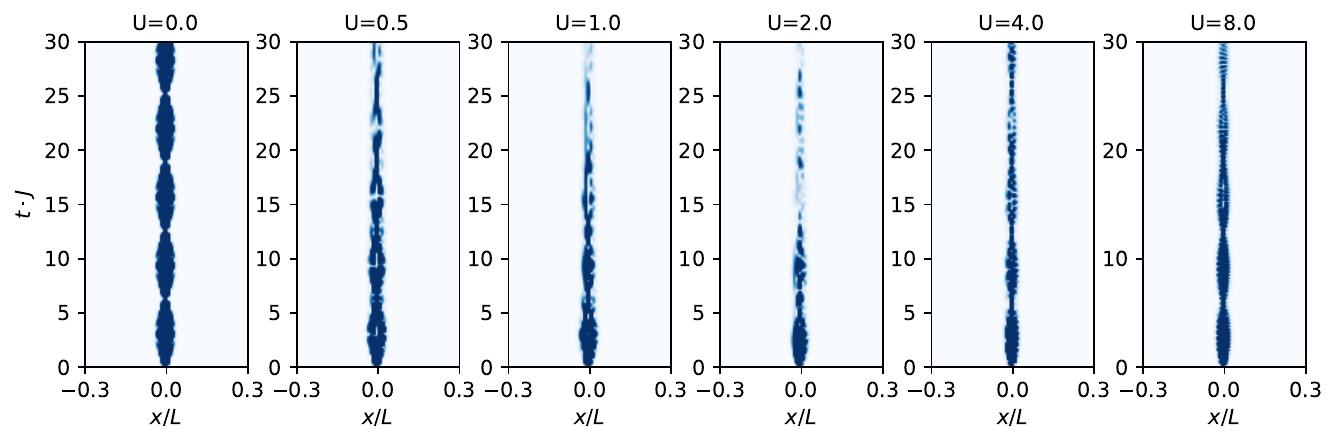}
	 \caption{
	Absolute value of the doublonic Green’s function, $|G_d(x,t)|$, shown for various interaction strengths $U$ as labeled in each panel, in the 
	presence of an external electric field $E = 1$. At $U = 0$, clear periodic Bloch oscillations are observed with period $T_B = 2\pi/\omega_B$. 
	Remarkably, this periodicity persists even for finite interactions $U > 0$. We adopt the vector potential gauge. }
	 \label{fig:G_x_t}
	\end{center}
\end{figure*}

\begin{figure}[htb!]
	\begin{center}
	 \includegraphics[width=0.95\columnwidth]{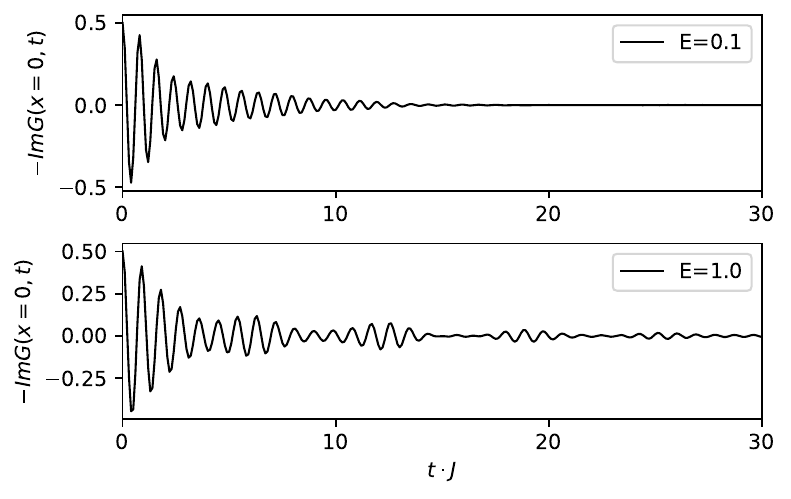}
	 \caption{
Time evolution of the local doublonic Green’s function, $G_d(x=0,t)$, evaluated at the center of the chain for fixed interaction strength $U=8$ and two different electric field amplitudes $E$, as indicated in each panel. Calculations done using the vector potential gauge. 
}
	 \label{fig:G_x_0_t}
	\end{center}
\end{figure}

\begin{figure*}[htb!]
	\begin{center}
	 \includegraphics[width=1.9\columnwidth]{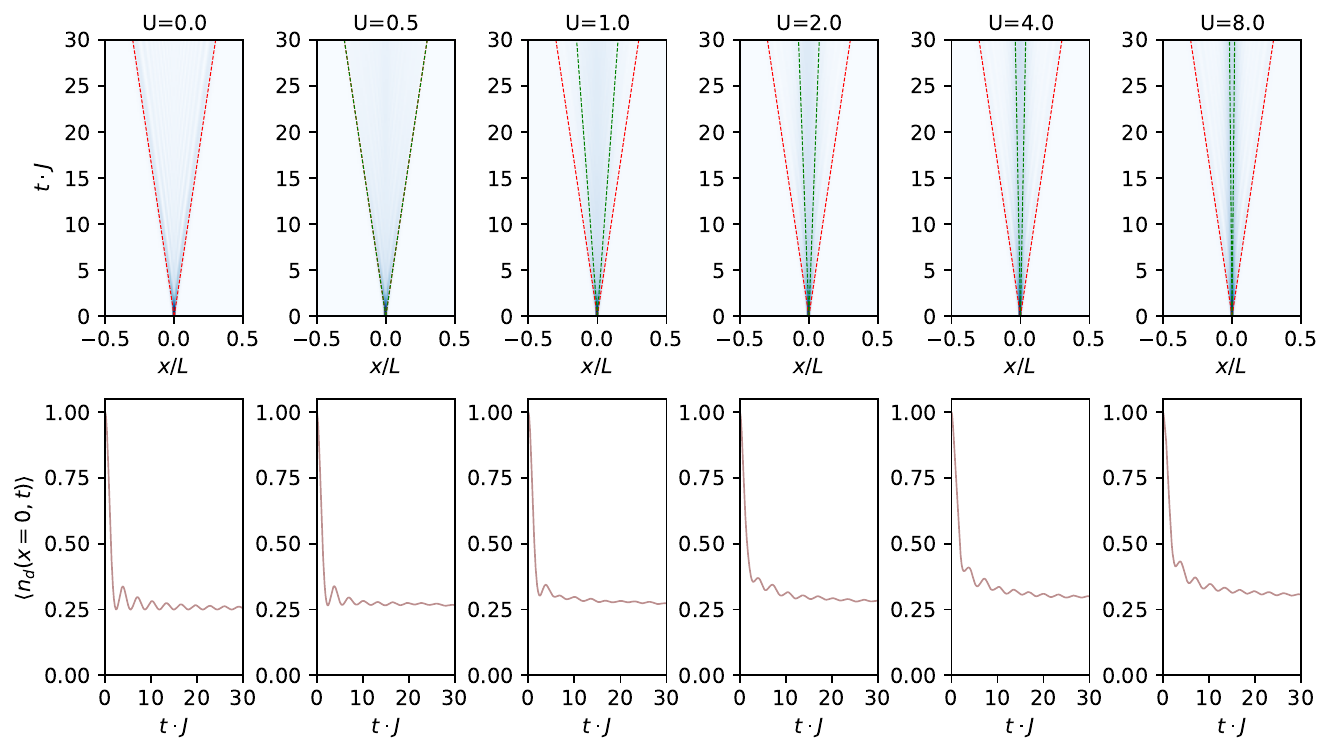}
	 \caption{
Top row: Time evolution of the doublon occupation $n_d(x,t)$, initially injected at 
the center of the chain, for various interaction strengths U (as labeled in each 
panel). The data show the formation of a light cone, indicating ballistic spreading 
of doublons with a well-defined velocity.
The dashed red lines mark the light cone corresponding to single-particle excitations, which propagate with a characteristic velocity $v_F \propto J$. In contrast, the green lines highlight the doublon light cone, indicating propagation at a reduced velocity $v_d \propto J^2/U$, consistent with the effective doublon dynamics at strong coupling.
Bottom row: Decay of $\average{n_d(x=0,t)}$ in the center of the chain  as a function of time. 
}
	 \label{fig:n_x_t}
	\end{center}
\end{figure*}

\begin{figure*}[htb!]
	\begin{center}
	 \includegraphics[width=1.9\columnwidth]{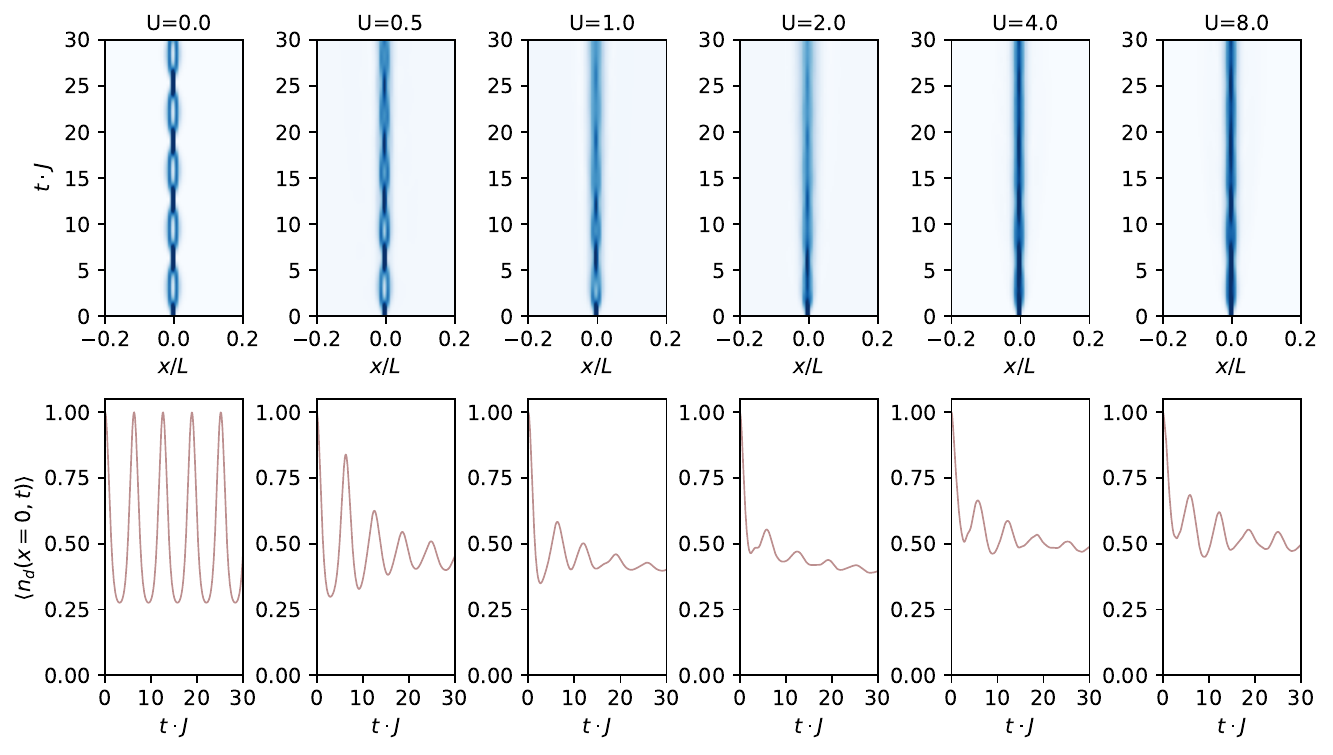}
	 \caption{
	Time evolution of the doublon occupation $n_d(x,t)$, initially localized at the center of the chain, is shown for various interaction strengths $U$ (as indicated in each panel) under a uniform electric field $E = 1$. For $U = 0$, clear Bloch oscillations appear, with a period given by $T_B = 2\pi/\omega_B$, characteristic of coherent dynamics in a static field. The bottom row displays the time dependence of the doublonic occupation $\average{n_d(x=0,t)}$ in the center of the chain. In the non-interacting case ($U = 0$), the doublon number exhibits undamped oscillations, while increasing $U$ leads to interaction-induced dephasing and decay. Here we used the 
	scalar potential gauge. 
}
	 \label{fig:n_E_x_t}
	\end{center}
\end{figure*}

We now turn to the interacting regime with $U > 0$. The top panels of Fig.~\ref{fig:Ad_k_w} illustrate the evolution of 
the momentum-resolved spectral function $A_d(k,\omega)$ as the interaction strength is progressively increased. In the 
non-interacting limit ($U=0$), particle-hole symmetry ensures that the two-hole ($2\text{-}h$) and two-particle ($2\text{-}
p$) contributions are identical, resulting in a single symmetric band structure. However, as $U$ increases, this symmetry 
is broken, and the spectral weight splits into two well-defined bands, separated by $\sim 2U$, corresponding to 
distinct doublon addition and removal processes. For moderate interaction strengths ($U \lesssim 2\text{–}3$), the bands 
still partially overlap due to their finite width. In contrast, at larger values of $U$, the bands become fully separated: 
the lower-frequency region ($\omega < 0$) is dominated by $2\text{-}h$ excitations, while the upper-frequency region 
($\omega > 0$) is primarily associated with $2\text{-}p$ processes. This band separation  highlights the role of interaction in reshaping the excitation spectrum.

We now examine the combined effect of interactions and an external electric field by considering finite values of both 
$U$  and $E$. The top panels of Fig.~\ref{fig:Ad_E_k_w} display the momentum-resolved doublonic spectral function $A_d(k,
\omega)$ for several values of $U$ in the presence of a fixed electric field $E=1$. As discussed before, at $U=0$, the 
spectral function exhibits the characteristic Wannier-Stark ladder: a series of sharp resonances located at integer 
multiples of the Bloch frequency $\omega_B$, as dictated by Eq.~\eqref{eq:WSL}. When interactions are introduced, the 
spectrum evolves significantly: the resonant structure persists, but the spectral weight begins to split, revealing the 
emergence of two distinct excitation branches associated with $2\text{-}h$ and $2\text{-}p$ processes. As $U$ increases, 
these bands shift away from each other by an energy approximately equal to $2U$, consistent with the interaction-induced 
doublon dynamics observed in the $E=0$ case. Notably, even for large interaction strengths, the Wannier-Stark resonances 
remain visible within each band, still they get damped, indicating that the electric field continues to induce coherent ladder-like structures within the otherwise interaction-split spectrum if the electric field is strong enough. 

\subsection{Time dependent Bloch oscillations}

To further investigate Bloch oscillations~\cite{Zhang.2024}, we compute the space- and time-resolved doublon Green’s function, $G_d(x,t)$, as defined in Eq.~\eqref{eq:G_d_vec}. The amplitude $|G_d(x,t)|$ is shown in Fig.~\ref{fig:G_x_t} for several interaction strengths $U$, in the presence of an external electric field $E = 1$. The results clearly reveal that doublonic excitations undergo oscillatory motion with a frequency set by the Bloch frequency $\omega_B$. 
This translated into a periodic time dependence with period $T_B = 2\pi/\omega_B$.
This behavior is particularly transparent in our simulations at infinite temperature, 
as the absence of initial entanglement allows for a clean  
observation of the oscillatory dynamics.

Interestingly, our simulations reveal that the Bloch oscillations of doublons persist 
even in the presence of strong interactions $U > 0$. Although the interactions renormalize the doublon dispersion and affect 
their internal structure, they do not completely destroy the coherence of the oscillatory motion. As a result, the doublonic Green’s 
function $G_d(x,t)$ retains its Bloch periodicity across a wide range of interaction strength.

To further elucidate the doublon dynamics, we present in Fig.~\ref{fig:G_x_0_t} the time evolution of the local doublonic Green’s function $G_d(x=0,t)$, computed at the center of the chain for a representative strong interaction strength $U = 8$.
The signal exhibits a complex structure: rapid oscillations with a 
characteristic frequency $\omega_U \approx U$, associated with high-energy doublonic 
dynamics, are modulated by slower oscillations at the Bloch frequency $\omega_B $, 
resulting from the external electric field. Importantly, the amplitude of the signal 
decays exponentially over time, reflecting the loss of doublon coherence due to the 
interplay of strong interactions and field-induced dephasing. The overall behavior is 
well captured, up to an overall prefactor, by the expression
\begin{equation}
-\mathrm{Im}\, G_d(x=0, t) \propto  \cos(\omega_U t)\cos(\omega_B t) e^{-\gamma t},
\end{equation}
where $\gamma$ is a decay rate that depends sensitively on both $U$ and $E$. For weak 
electric fields, the small Bloch frequency $\omega_B$ is effectively obscured by the fast decay, while for sufficiently strong fields, the modulation at $\omega_B$ becomes 
clearly visible even at large $U$. This damped oscillatory behavior, when Fourier 
transformed, gives rise to the broadened Stark ladder structure seen in the spectral 
function (Fig.~\ref{fig:Ad_E_k_w}), confirming that coherent Bloch oscillations of 
doublons can survive in the strongly correlated regime when the electric field is large 
enough.

\subsection{Doublon quantum walk}

We further explore this behavior by examining the spatial and temporal evolution of the doublon occupation $\average{n_d(x,t)} = \average{d^\dagger_x(t) d_x(t)}$, following the injection of a localized doublon into an infinite-temperature background. As shown in Fig.
\ref{fig:n_x_t}, the propagation pattern reveals two distinct light-cone structures. The outer light cone, largely independent 
of interaction strength, corresponds to fast-moving single-particle excitations with velocity $v_F \propto J$. In contrast, the 
inner light cone—emerging only for finite $U$—reflects the slower dynamics of bound doublons, with a velocity $v_d \propto J^2/
U$ determined by the effective second-order hopping amplitude~\cite{sensarma2010dynamics, ostahie2023multiparticle}.
This separation of time and length scales highlights the coexistence of incoherent single-particle spreading and coherent 
doublonic motion, and emphasizes the role of interactions in stabilizing doublons as emerging quasiparticles. 

In Fig.~\ref{fig:n_E_x_t} we present results for the occupation of a doublon when a 
 finite electric field $E \neq 0$ is applied. In the non-interacting limit ($U = 
0$), the system displays coherent Bloch oscillations in the doublon density with period $T_B$~\cite{Davies.1988}. These oscillations 
persist indefinitely in time, indicating robust coherence. As $U$ is increased, the oscillations become damped, 
signifying that the combined effect of the field and interactions acts as a drive that induces doublon decay via coupling to multi-particle scattering 
channels.

\section{Conclusion}

In this work, we explored the spectral and dynamical properties of doublons in the one-dimensional Hubbard model at infinite temperature, with a 
particular focus on their behavior under a static electric field. Using the Chebyshev polynomial expansion in the superfermionic formalism, we 
computed high-resolution doublonic spectral functions $A_d(k,\omega)$ and analyzed their evolution with interaction strength $U$ and electric 
field $E$.

At $U=0$, and finite $E$, the doublonic spectral function exhibits the well-known Wannier-Stark ladder structure, with sharp resonances at integer multiples of 
the Bloch frequency $\omega_B$, clearly reflecting the coherent dynamics of charge carriers in a tilted potential. As interactions are gradually 
introduced, the spectrum splits into distinct two-hole and two-particle bands, separated by an energy scale proportional to $\sim 2U$. Remarkably, the 
Wannier-Stark structure persists even at strong interactions, indicating that doublons remain coherent and well-defined quasiparticles.
We further examined the real-time propagation of doublons via the time-dependent Green's function, revealing robust and long-lived 
Bloch oscillations. These oscillations persist over a broad range of $U$, showcasing the resilience of doublonic coherence even in the presence 
of strong correlations. Our results provide a unified picture for doublonic spectral properties and their dynamical behavior, and highlight 
the utility of the Chebyshev-superfermion framework for studying correlated systems.
When a doublon is injected in the infinite-temperature background it reveals a rich interplay between interaction 
strength and external driving. At zero electric field, the propagation of doublonic excitations exhibits a 
two-fold light-cone structure, and despite not being a conserved quantity, 
the total doublon occupation shows emergent conservation behavior in the strong-coupling regime due to the suppression 
of kinetic processes. Introducing a finite electric field leads to coherent Bloch 
oscillations in the doublon density for $U=0$, which decays at finite interactions. 

\begin{acknowledgments}
This work received financial support from CNCS/CCCDI-UEFISCDI, under projects number 
PN-IV-P1-PCE-2023-0159 and PN-IV-P1-PCE-2023-0987. We acknowledge the Digital 
Government Development and Project Management Ltd.~for awarding us access to the Komondor 
HPC facility based in Hungary.

\end{acknowledgments}

\bibliography{references}

\end{document}